\documentclass[11pt]{article}

\usepackage[utf8]{inputenc}
\usepackage[T1]{fontenc}
\usepackage{amsmath}
\usepackage{amsfonts}
\usepackage{mathtools}
\usepackage[dvipsnames,svgnames,x11names]{xcolor}
\usepackage{graphicx}
\usepackage{tikz}
\usepackage{url}
\usepackage{verbatim}
\usepackage[normalem]{ulem}
\usepackage{array}
\usepackage{xspace}
\usepackage{mathpazo}
\usepackage{float}
\usepackage{graphicx}
\usepackage{wrapfig}
\usepackage{caption}
\usepackage{cite}
\usepackage[top=1.00in,bottom=1.00in,left=1.00in,right=1.0in]{geometry}
\usepackage{todonotes} %[disable]

\title{Insights into human evolution from large genetic biobanks}

\author{
Jeffrey P. Spence$^{1,2,\ast}$ and Roshni A. Patel$^{3,4}$\\
\small $^1$ Institute for Human Genetics, University of California, San Francisco\\
\small $^2$ Department of Epidemiology \& Biostatistics,  University of California, San Francisco\\
\small $^3$ Department of Data Science, University of Oregon\\
\small $^4$ Institute for Ecology and Evolution, University of Oregon\\
\small $^\ast$ Correspondence to: \texttt{jeff.spence@ucsf.edu} \\
}
\date{}
\begin{document}
\maketitle

%TC:ignore
\begin{abstract}
The development of large genetic biobanks with deep phenotyping and whole-exome or whole-genome sequencing is enabling an ever-deeper understanding of human evolution.  Here we review recent developments in two main research directions: 1) using the increasing sample sizes to infer extremely strong evolutionary constraint and 2) leveraging biobanks' extensive phenotyping to relate the effects of natural selection on variants to their impacts on traits.  In particular, there has been recent progress in gene-specific estimates of the strength of selection acting against loss-of-function mutations, as well as growing evidence supporting the importance of pleiotropic stabilizing selection on traits in shaping patterns of genetic diversity. We also discuss how these findings are, in turn, improving our understanding and interpretation of the genetic associations discovered in biobanks.  Throughout, we outline open questions and promising directions for future research.
\end{abstract}
%TC:endignore

\section*{Introduction}
Genetic biobanks \cite{ziyatdinov2023genotyping,all2024genomic,uk2025whole} are an unprecedented resource for the study of human genetics, generating ever more associations between variants and traits that explain increasing amounts of heritability \cite{yengo2022saturated,wainschtein2026estimation}. Beyond finding associations, biobanks have also provided new insights into human evolution. Here we review progress in estimating evolutionary constraint and understanding how selection on variants is mediated by their effects on traits, but biobanks have enabled progress more broadly.

Biobanks differ from earlier population genetics datasets along several axes. Population genetics datasets typically range in sample sizes from the hundreds \cite{mallick2016simons} to about a thousand \cite{10002012integrated,bergstrom2020insights}. In contrast, the UK Biobank (UKB) contains half a \emph{million} individuals \cite{uk2025whole}.   The gnomAD consortium has aggregated allele frequencies across the protein-coding genome for over \emph{700{,}000} individuals \cite{guez2026integrating}.  

Earlier population genetics datasets favored sampling broadly to better capture diversity, which can be extremely useful for learning demographic history \cite{mallick2016simons,bergstrom2020insights}.  On the other hand, existing biobanks are notoriously unrepresentative of global diversity \cite{mills2019scientometric}.  Yet, in terms of absolute numbers, biobanks still represent the largest source of data. For example, about 3/4 of gnomAD consists of individuals labeled by the consortium as having ``European ancestries'' but it still contains data from over \emph{35 thousand} individuals labeled as ``African/African American'' \cite{gnomad_stats}.  Furthermore, while diverse samples increase power for common variants \cite{wojcik2019genetic} and enable analyses of heterogeneous environments and genetic backgrounds (e.g., \cite{patel2022genetic}), most genetic variation and all fundamental aspects of biology are shared across groups \cite{biddanda2020variant,stolyarova2025distribution}.

\begin{figure}
    \centering
    \includegraphics[width=0.5\textwidth]{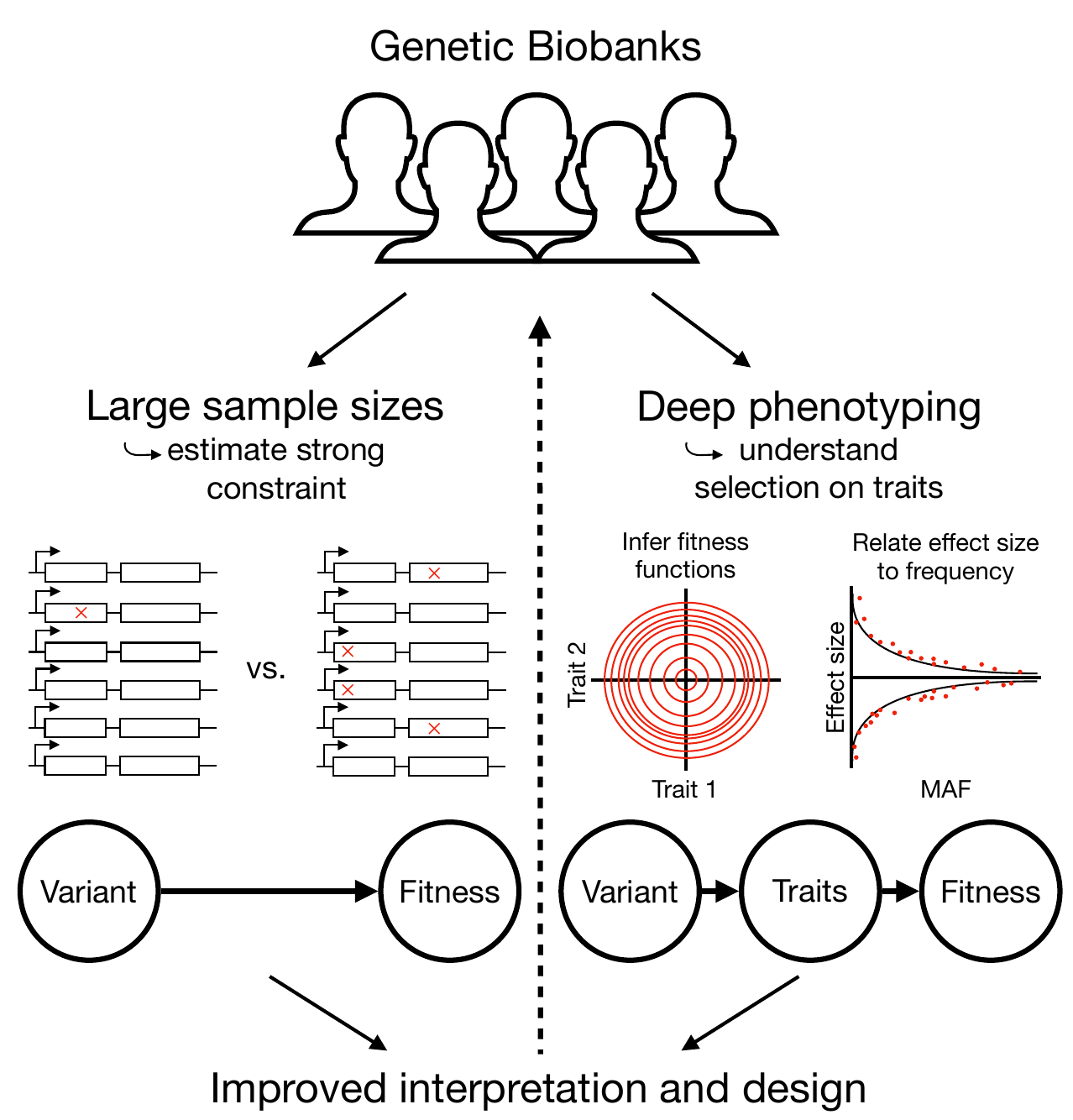}
    \caption{\textbf{Insights into human evolution from biobanks}\\ Genetic biobanks provide extremely large sample sizes and deep phenotyping.  Large sample sizes enable inference of strong natural selection, and extensive phenotyping enables learning how selection acts on traits to shape genetic diversity.  These evolutionary insights then improve the interpretation and design of association studies in biobanks.
    }
    \label{fig:fig1}
\end{figure}

Previously, population genetics datasets were unique in using whole-genome sequencing to obtain unbiased estimates of genetic variation.  Biobanks traditionally used genotyping arrays and imputation, which bias the frequency spectrum. That has changed, and whole-exome or whole-genome sequencing is becoming the norm \cite{uk2025whole}.

Beyond sample size, the other major advantage of biobanks is, obviously, the phenotypes. UKB has thousands of phenotypes from anthropometric measurements to serum proteomics \cite{sun2023plasma,uk2025whole}. Combining phenotypes with measures of fitness provides a path for understanding \emph{why} variants are constrained in terms of their impacts on traits.

All together, biobanks now represent a massive increase in the amount of data that can be used to understand human evolution. Here we outline recent progress in this direction as well as how our understanding of human evolution has, in turn, improved our interpretation of the genetic associations found in biobanks (Figure~\ref{fig:fig1}).

\section*{Estimating evolutionary constraint from large sample sizes}

One goal of human population genetics is to estimate how natural selection is acting on every variant in the genome.  Recent work toward this goal has focused on purifying selection, which reduces the frequency of deleterious alleles.  Evolutionary constraint can then be estimated by measuring the depletion of genetic diversity relative to a neutral expectation. The large samples provided by biobanks enable more precise estimates of the frequency of ultra-rare variants, increasing power \cite{spence2023scaling}.

\begin{figure}
    \centering
    \includegraphics[width=\textwidth]{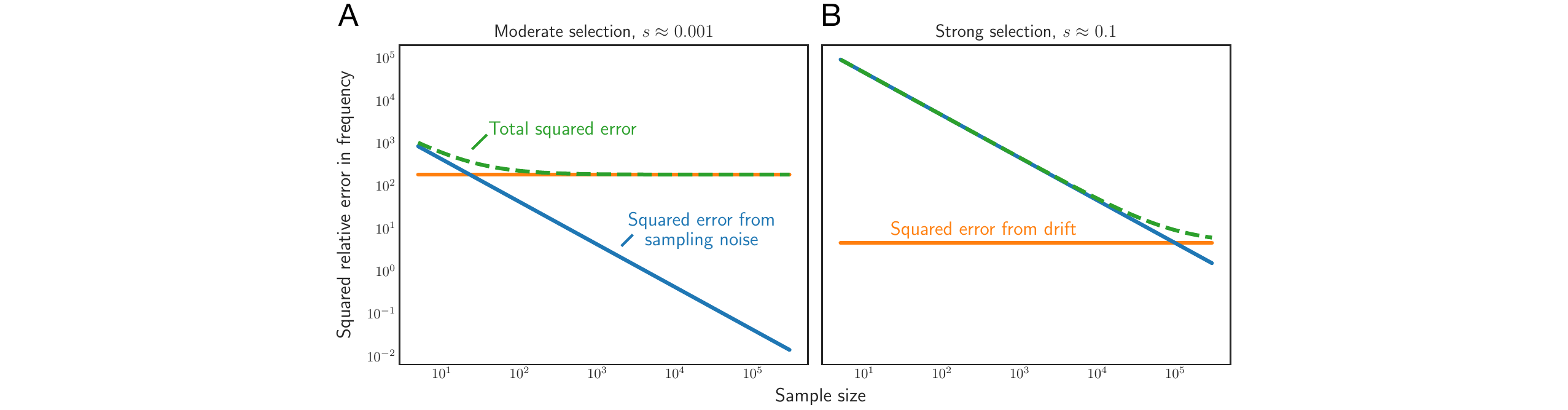}
    \caption{\textbf{Diminishing returns for estimating constraint from increasing sample sizes}\\ Panels A and B decompose the error in the sample allele frequency into the component due to sampling error and the component due to drift.  Specifically, we assume the population frequency, $f$, is drawn according to the Discrete-time Wright--Fisher model assuming an out-of-Africa demographic model (called the CEU model in \cite{spence2023scaling}), a CpG-like mutation rate of $1.25\times10^{-7}$, and a selection coefficient of either 0.00115 (panel \textbf{A}) or 0.105 (panel \textbf{B}).  We then assume that --- given the population frequency --- the sample frequency, $\widehat{f}$, in a sample of size $n$ is $1/2n$ times a draw from a $\text{Binomial}(2n, f)$ distribution. The total squared relative error (green lines) is then $\mathbb{E}[((\widehat{f} - \mathbb{E}[f]) / \mathbb{E}[f])^2]$.  This can be decomposed into a component due to sampling noise that decreases with sample size : $\mathbb{E}[((\widehat{f}-f)/\mathbb{E}[f])^2]$  (blue lines), and a component due to genetic drift that is constant across sample sizes: $\mathbb{E}[((f-\mathbb{E}[f])/\mathbb{E}[f])^2]$ (orange lines).  Even for extremely strongly selected alleles, biobanks have already reached sample sizes ($n\approx200{,}000$) where additional samples provide essentially no information about constraint.
    }
    \label{fig:fig2}
\end{figure}

Yet, even with these huge sample sizes, it is incredibly difficult to estimate the strength of selection acting on a single variant.  Larger sample sizes decrease sampling noise, allowing for more accurate estimation of the population frequency. Yet, even with a \emph{perfect} estimate of the population frequency, the evolutionary noise caused by drift would remain, resulting in severely diminishing returns as sample sizes increase (Figure~\ref{fig:fig2}) \cite{spence2023scaling}. To average out the effects of drift, approaches must pool information across variants, ideally variants that are equally constrained.  

Loss-of-function mutations (LoFs) are particularly useful in this regard.  There are many different possible unique LoF alleles for any given gene, but they all effectively eliminate the copy of the gene in which they reside, essentially having the same effect on fitness (but see \cite{blakes2024regional}).  Many approaches pool information across LoFs within a gene, including LOEUF \cite{guez2026integrating} and related metrics \cite{lapolice2023unsupervised}, and various estimates of the strength of selection against heterozygous LoFs (e.g., most recently \cite{agarwal2023relating,zeng2024bayesian}).

Across these lines of research, LoFs have been found to range from being effectively neutral to having severe fitness consequences \cite{agarwal2023relating,zeng2024bayesian}. Constraint is strongly predicted by gene features including regulatory complexity (e.g., number of distinct isoforms), expression patterns across tissues, and various gene sets (e.g., olfactory receptors are less constrained) \cite{zeng2024bayesian}. LoFs in constrained genes also tend to be associated with larger effects on traits in the UKB \cite{spence2026specificity}.  Together, these results are just beginning to illuminate which biological processes are constrained by evolution.

To move beyond LoFs, a recent method, \texttt{Gnocchi}, pools variants into 1kb windows to estimate regional levels of constraint \cite{chen2024genomic}.  While this approach suffers from the heterogeneity of effects within each window --- not all variants within any window should be constrained to the same extent --- the results are promising and revealed insights into the biological targets of constraint, including an unexpected enrichment of constraint at loci encoding microRNAs \cite{chen2024genomic}.  Approaches have also pooled information at the level of individual amino acids \cite{zhang2024genetic} or in terms of 3D space as determined by a protein's structure \cite{bajracharya2025fine}.

Another approach to pool information is GeneBayes \cite{zeng2024bayesian}, which uses empirical Bayes to learn a mapping from a gene's features (e.g., expression patterns across tissues) to a Bayesian prior over the strength of selection against heterozygous LoFs for that gene. On an intuitive level, this provides a form of ``soft pooling'', where genes with similar features have similar prior distributions. Extending this approach to use information at the \emph{variant} level (e.g., from deep mutational scanning data \cite{beltran2025site} or protein/genomic language models \cite{brandes2023genome,ye2025predicting}), may allow for accurate estimates of base pair-level constraint exome- or genome-wide.

High mutation rates can also be leveraged to estimate constraint. For example, Agarwal and Przeworski showed that the mutation rate for methylated CpG dinucleotides is high enough that one is almost guaranteed to find some individual carrying a TpG allele in biobank-scale datasets --- provided that the variant is neutral \cite{agarwal2021mutation}.  Therefore, if a methylated CpG site is not segregating in a biobank, it is strong evidence of constraint. 
Similar ideas have been or could be applied to other high mutation rate variants, including at short tandem repeats \cite{huang2025genome,fernandez-luna2026statistical} or highly-expressed non-coding genes \cite{thornlow2018transfer,seplyarskiy2023mutation,chen2024novo}.

Regardless of methodological choices, it is important to carefully operationalize ``constraint''.  Some approaches, like LOEUF \cite{guez2026integrating}, use \emph{ad hoc} statistics that capture the intuition that variation should be depleted in constrained regions, but are correlated with technical confounders including gene length and sample size \cite{fuller2019measuring}.  This hinders comparison of these statistics across genes or cohorts, or to other \emph{ad hoc} statistics.  As such, Agarwal and colleagues argued for the use of explicit population genetics models, where evolutionary constraint can be formulated as interpretable selection coefficients \cite{agarwal2023relating}.

There are several open questions in this space.  First, how consistent is selection across space and time?  Work comparing LoF frequencies across ancestry groups suggests that at least in aggregate, selection is largely similar across different ancestry groups \cite{stolyarova2025distribution}, but it is unclear whether this holds for any given gene or for more weakly selected variants.  Less is known about the similarity of selection against LoFs in orthologous genes across species.  GeneBayes was recently used to estimate LoF constraint across mice, fruit flies, and yeast, finding that similar features were predictive of constraint across species, but again this is in aggregate across genes \cite{burcin2026gene}.

Another area of active research is the role of recessive selection.  Most approaches discussed above are only well-powered to estimate selection in heterozygotes \cite{fuller2019measuring}, as for most genes --- even recessive disease genes --- most of the effect on LoF allele frequency is driven by selection occurring in hetozygotes \cite{judd2025allele}. Aggregating information across genes has been shown to be a promising approach for estimating recessive effects \cite{balick2022overcoming}, but methods for data-driven pooling are needed.

\section*{Understanding how natural selection acts on traits}

Even with our expanding understanding of which regions of the genome are constrained, it is unclear \emph{why} those regions are constrained. Biobanks promise to provide a partial answer to this question by relating variants' effects on traits to their fitness consequences.

It has long been appreciated that there is selection against trait-affecting variants in humans. Early work characterized the relationship between a variant's frequency and its magnitude of effect on a trait of interest (e.g., \cite{zeng2018signatures}).  Across many traits, strong negative relationships were found, suggesting that variants with larger effects on traits are more constrained \cite{koch2021maintenance}.  Yet, these approaches were purely phenomenological and hence it was unclear what types of natural selection on traits, if any, would generate the observed relationships.  

Several lines of evidence now support the idea that many traits are shaped by pleiotropic stabilizing selection, and that this can explain observed relationships between allele frequency and trait effects \cite{sella2019thinking}. Early work estimated the relationship between traits and fitness by directly regressing individuals' lifetime reproductive success against functions of phenotypes \cite{sanjak2018evidence}.  More recent work leveraged the fact that the effect of stabilizing selection on an allele is independent of the sign of its effect and compared the relationship between allele frequency and effect size for variants that have positive and negative effects on traits \cite{koch2024genetic}.  Another approach looked at GWAS hits, and to avoid biases due to ascertainment, compared the frequency of variants in one population conditioned on their frequency in the GWAS population to expectations under various models of selection \cite{patel2025characterizing}. Finally, the LD score regression approach was extended to identify whether there tends to be positive LD between variants with opposite effects, a pattern expected under stabilizing selection \cite{zhang2023pervasive}.  All of these disparate approaches found that for many traits the evidence better support stabilizing selection than other models of selection. 

It appears that stabilizing selection even provides a good model for disease risk for many diseases, even though one might instead assume that selection should directionally act to reduce disease risk. Indeed, Berg and colleagues developed a model of the relationship between effect sizes and frequency under directional selection on risk, and found poor fit to data from the UKB, instead finding a better fit to pleiotropic stabilizing selection \cite{berg2025mutation}.  

Stabilizing selection also provides a good fit to the genetic architectures of complex traits \cite{simons2025simple}.  With only a handful of parameters (a trait's heritability, mutational target size, and a distribution of selection coefficients shared across traits), Simons and colleagues were able to accurately predict the distribution of $p$-values and allele ages for GWAS hits from held-out loci \cite{simons2025simple}.  In the original work, it was found that a single distribution of selection coefficients provided a good approximation across many disparate traits.  This similarity of genetic architectures may be due in part to emergent properties from the organization of gene regulatory networks \cite{aguirre2025gene,aguirre2026regulatory}.  More recently, it was shown that inference can be improved by instead assuming a shared distribution of selection coefficients for traits that are mediated by the same tissue (e.g., traits primarily driven by the central nervous system) \cite{zhu2026genetic}.  This is potentially consistent with the omnigenic model \cite{boyle2017expanded}, where most genes active within a given cell type contribute to any trait mediated by that cell type.

There are a number of open questions in the study of how stabilizing selection shapes human traits.  Recent theoretical work has just begun exploring the effect of shifting trait optima \cite{hayward2022polygenic,milligan2025when,bertram2026strong}; weak LD between unlinked variants \cite{negm2026effect}; population structure \cite{veller2024stabilizing,ragsdale2025archaic,li2026effect}; and background selection \cite{li2026background}. In contrast to neutral models, models of stabilizing selection appear to be much more sensitive to these sorts of fine-scale aspects of the evolutionary model.  As such, more work is needed to understand the extent to which our intuition from panmictic populations at equilibrium is relevant to humans.

Furthermore, the exact relationship between traits and fitness remains unclear.  That is, what combination of traits does selection act on, and to what extent?  It is particularly challenging to disentangle which traits are directly selected upon as opposed to merely being genetically correlated with a target of selection.  Here, the deep phenotyping provided by biobanks will be particularly useful.

Another interesting direction would be to bridge this line of work on stabilizing selection with the work on evolutionary constraint described in the previous section.  Is most ``evolutionary constraint'' really just stabilizing selection acting against new trait-altering alleles?  There is conflicting evidence in these directions.  We recently showed that evolutionary constraint is strongly predictive of the magnitude of LoF effects on traits in the UKB, consistent with stabilizing selection \cite{spence2026specificity}.  Yet, in another line of work, we found that, when averaged across genes, LoFs and whole-gene duplications often affect traits in the \emph{same} direction \cite{milind2026buffering}.  That is, large effect mutations are biased to affect the trait in a particular direction.  Some of this could be induced by measuring phenotypes on the ``wrong'' scale \cite{cole2026representation}, but the bias observed in LoFs and duplications is usually in the direction associated with accelerated aging and other negative outcomes \cite{milind2026buffering}.  Overall, this suggests that there may be other mechanisms by which extremely large effect mutations affect fitness.

Finally, here we focused on approaches assuming mutation--selection--drift balance, but this is separate from the question of whether polygenic adaptation has occurred in the course of human history.  Detecting polygenic adaptation is theoretically challenging as it occurs astonishingly quickly via very small frequency changes at many loci \cite{hayward2022polygenic}.  One approach uses cross-population allele frequency measures \cite{berg2014population,chen2020evidence} and aggregates signal across many trait-associated variants. Unfortunately, this results in technical challenges as even very weak confounding in GWAS estimates of trait associations can result in spurious findings \cite{berg2019reduced,sohail2019polygenic}, but there has been recent promising work on developing methods robust to this confounding \cite{blanc2025testing}.

\section*{Implications of evolution for understanding biobanks}

\begin{figure}
    \centering
    \includegraphics[width=\textwidth]{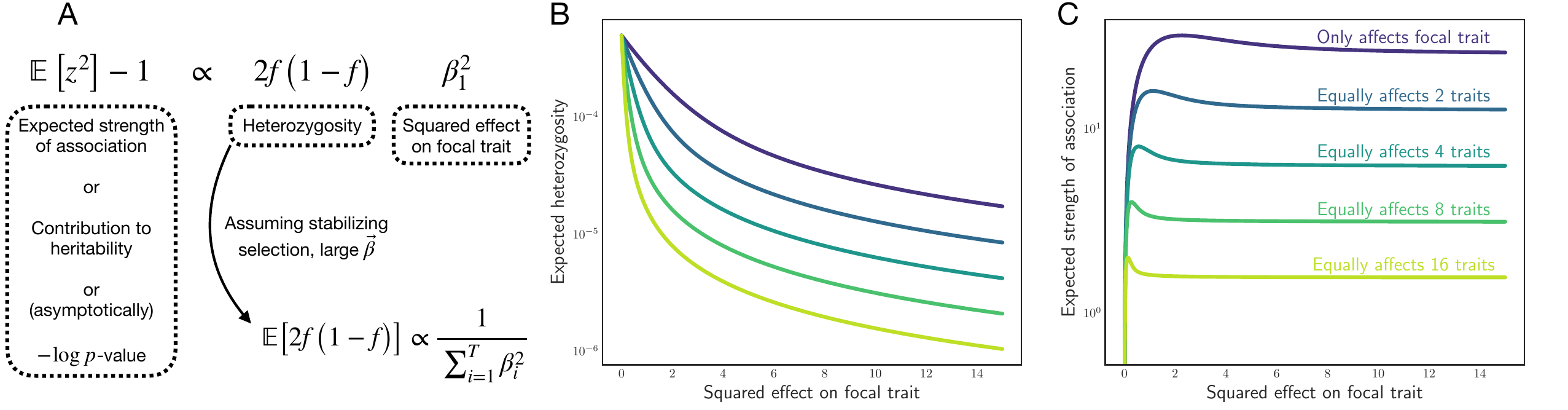}
    \caption{\textbf{Effects of pleiotropic stabilizing selection on association studies}\\ \textbf{A}) The expected strength of association in association studies, including GWAS, is proportional to the heterozygosity times the squared effect on the focal trait.  This is also proportional to the variant's (genic, additive) contribution to heritability, and for strong associations is proportional to the $-\log p\text{-value}$ \cite{spence2026specificity}.  Under stabilizing selection, the expected heterozygosity at sites with large effects is proportional to the inverse of the squared effects on all fitness relevant traits, assuming a particular choice of trait coordinates \cite{spence2026specificity}.  \textbf{B}) Graphical representation of the relationship between expected heterozygosity and squared effect size under mutation--selection--drift balance assuming the population-scaled mutation rate, $4N_e\mu$, is $0.0005$, and the squared effect sizes are measured in units of the population-scaled selection coefficient (see \cite{simons2018population} for details). \textbf{C}) Multiplying the curves in \textbf{B} by the squared effect size on the focal trait (and a constant of proportionality) gives the expected strength of association (equivalently their contribution to heritability).
    }
    \label{fig:fig3}
\end{figure}

While biobanks have been tremendously useful in learning about constraint and selection on traits, this deeper understanding of human evolution is also proving to be useful for interpreting GWAS and other association tests in biobanks.  Power in association studies increases with both effect size (how much a variant affects the trait) and minor allele frequency (Figure~\ref{fig:fig3}\textbf{A}), but stabilizing selection acts to make the most impactful variants the rarest.  This simple observation has had wide-ranging implications for the interpretation of association studies.

At the broadest scale, natural selection explains many of the features of the genetic architectures of complex traits. Stabilizing selection causes effect sizes and allele frequencies to be negatively correlated (Figure~\ref{fig:fig3}\textbf{B}), causing variants with vastly different effect sizes to contribute roughly equally to heritability (Figure~\ref{fig:fig3}\textbf{C}), a phenomenon referred to as ``flattening'' \cite{simons2018population,oconnor2019extreme}.  Flattening explains the extreme polygenicity seen for many traits (but see \cite{oconnor2026principled} for a discussion of the technical difficulties of rigorously defining ``polygenicity'').  Indeed, there are now over 12,000 independent genome-wide significant loci for height, but they collectively do not explain all of the heritability \cite{yengo2022saturated}.  It is not until variants within 35kb on either side of any of these genome-wide significant loci are included (covering ${\approx}22$\% of the genome) that all of the heritability can be recovered \cite{yengo2022saturated}.  On the one hand, this says something about the \emph{biology} of height --- variants must have some effect on height in order to contribute to heritability, and so an astounding number of variants must affect height to some extent.  Without flattening, however, heritability would be more strongly concentrated in a few common variants of large effect.  Height is not unique: heritability is spread across large numbers of variants across many traits \cite{simons2025simple,oconnor2026principled}. 

Stabilizing selection also implies that association studies are better powered to find trait-specific variants \cite{spence2026specificity}.  Under pleiotropic stabilizing selection, the frequency of a variant is determined by its effects across \emph{all} fitness-relevant traits (Figure~\ref{fig:fig3}\textbf{A}), so affecting other traits reduces a variant's frequency (Figure~\ref{fig:fig3}\textbf{B}), reducing power (Figure~\ref{fig:fig3}\textbf{C}).  This explains observed differences in which genes are discovered by association tests when considering rare protein-coding variants (burden tests) versus common non-coding variants (GWAS) \cite{weiner2023polygenic}.  In particular, coding variants affect a gene's product wherever it is expressed, whereas non-coding variants can be more context specific (e.g., only affecting expression of a gene in a particular cell type).  This allows GWAS to discover context-specific non-coding variants near highly pleiotropic genes, whereas coding variation in those same genes would be so constrained so as to be missed by burden tests.  Note that the difference in which genes are discovered using burden tests and GWAS is driven by natural selection-induced differences in power, not differences in the underlying biology of common and rare variants.  Indeed, when looking at the pathways affected by genes implicated by GWAS and burden tests there is much clearer overlap \cite{wright2026common,lamantrip2026unifying}.

A special case of the effect of trait specificity on power is that variants that are discovered to have an effect on gene expression (eQTLs) will be biased away from variants that also affect fitness-relevant traits, resulting in a lack of overlap between discovered eQTLs and discovered GWAS hits \cite{mostafavi2023systematic}. Again, this is driven by both biology and evolution.  Biology determines whether it is even possible for a variant to affect expression in a given cell type without affecting traits, but if such variants exist, then the effects of evolution will make them easier to discover as eQTLs. Larger eQTL sample sizes can overcome these differences in power, increasing the extent of overlap \cite{rosen2026higher}.  Recent work in domesticated animals supports the role of stabilizing selection, as eQTLs and GWAS hits have better overlap in these species where strong directional selection on agriculturally important traits masks the impacts of stabilizing selection \cite{connally2026farm}.

Interestingly, compared to the overlap between eQTLs and GWAS hits, chromatin accessibility QTLs appear to better overlap GWAS hits \cite{dudek2026meta}.  One possible explanation is that if a variant has a large impact on the expression of a highly trait-relevant gene, it will be strongly selected against, and hence will be difficult to discover as an eQTL.  Conversely, if it has a small impact on the expression of the gene, its small effect size will also make it difficult to discover as an eQTL.  In contrast, a variant might have a dramatic effect on chromatin accessibility at a weak enhancer, resulting in only moderate affects on expression, and hence be less exposed to selection \cite{dudek2026meta}.

Evolutionary insights promise to continue improving the interpretation and design of biobanks, and further work in this area is warranted.  There has been some promising work toward understanding how purifying selection interacts with sampling strategy in spatially-structured populations to impact GWAS power \cite{steiner2025study} and several studies on the impact of evolutionary forces on the portability of polygenic scores \cite{yair2022population,patel2025characterizing,anorve-garibay2025natural}, but there is significantly more work to be done understanding how biology interacts with evolution to shape the genetic associations we discover.

\section*{Discussion}

The recent development of enormous biobanks with whole genome sequencing and deep phenotyping has already provided new insights into human evolution.  These fantastic resources present an unprecedented opportunity, and there remain countless areas ripe for new research.  

At the same time, these huge sample sizes also present challenges for existing approaches.  Standard population genetic assumptions and approximations break down when samples are large \cite{spence2023scaling,schraiber2025estimation,guez2026integrating}, and methods must be computationally scalable.

Nevertheless, it is an exciting time to develop new methods, models, and theory that leverage  phenotypes and genotypes to better understand selection in terms of the traits on which it is acting, and how these effects percolate down to the dynamics of individual variants. 

%TC:ignore
\section*{Acknowledgments}
This work was funded in part by NIH grant R01HL175076.

\section*{Declaration of interest}
The authors declare no conflict of interest.

\bibliographystyle{myunsrt}
\bibliography{cogd}

%TC:endignore
\end{document}